\documentclass[prb,twocolumn,showpacs]{revtex4}
\usepackage{amsfonts,amsmath,mathrsfs,epsfig,amsbsy,bm,verbatim,subfigure}

\newcommand{\bra}[1]{\left\langle {#1} \right \vert}
\newcommand{\ket}[1]{\left\vert {#1} \right\rangle}
\newcommand{\expect}[1]{\langle {#1} \rangle}
\newcommand{\ketn}[1]{ {#1} \rangle}

\begin{document}

\title{Experimental and materials considerations for the topological superconducting state in electron and hole doped
 semiconductors: searching
for non-Abelian Majorana modes in 1D nanowires and 2D heterostructures}
\author{Jay D. Sau$^1$}
\thanks{Present Address: Department of Physics, Harvard University, Cambridge, MA 02138}
\author{Sumanta Tewari$^{2}$}
\author{S. Das Sarma$^1$}
\affiliation{$^1$Condensed Matter Theory Center and Joint Quantum Institute, Department of Physics, University of
Maryland, College Park, Maryland 20742-4111, USA\\
$^2$Department of Physics and Astronomy, Clemson University, Clemson, SC
29634}

\begin{abstract}
In proximity to an $s$-wave superconductor, a one- or two-dimensional, electron- or hole-doped semiconductor with a sizable spin-orbit coupling and a Zeeman splitting can
support a topological superconducting (TS) state. The semiconductor TS state
has Majorana fermions as localized zero-energy excitations at order parameter defects such as vortices and sample edges. Here
we examine the effects of quenched disorder from the semiconductor surface on the stability of the TS state in both electron- and hole-doped semiconductors.
By considering the interplay of broken time reversal symmetry (due to Zeeman splitting) and disorder we derive an expression for the disorder suppression of the superconducting quasiparticle gap in the TS state. We conclude that the effects of disorder can be
minimized by increasing the ratio of the spin-orbit energy with the Zeeman splitting. By giving explicit numbers we show that a stable TS state
is possible in both electron- and hole-doped semiconductors for experimentally realistic values
 of parameters.
We discuss possible suitable semiconductor materials which should be the leading candidates for the Majorana search in solid state
systems.
\end{abstract}

\maketitle


\section{Introduction}

In a recent paper \cite{sau-et-al} it has been shown that a 2D spin-orbit coupled electron-doped semiconductor, in proximity to a bulk $s$-wave
superconductor and an externally induced Zeeman splitting,
 can support a topological superconducting phase with Majorana fermion modes at vortex cores and sample edges i.e. at
order parameter defect locations. It has also been realized from dimensional reduction that the edge Majorana modes in a
 sample with rectangular geometry (of width $W$) turn into localized \emph{end} Majorana modes in the corresponding 1D geometry ($W\rightarrow 0$) and, further, in 1D the excitation gap above the Majorana modes (mini-gap) should scale as the induced superconducting gap $\Delta$ (thus avoiding the problem of a tiny mini-gap, $\Delta^2/\epsilon_F \sim 0.1$ mK, as in 2D $(p_x+ip_y)$-wave superconductors). The end Majorana modes in the 1D geometry should be experimentally visible in local zero-bias tunneling
  experiments. These results,
numerically confirmed soon thereafter, \cite{unpublished} have appeared
with analytical and numerical details for both 1D and 2D semiconductors in Ref.~[\onlinecite{long-PRB}].
  It has also been pointed out \cite{roman,oreg} that the required Zeeman splitting in 1D can be induced by a magnetic field parallel to the adjacent superconductor (thus avoiding the problem of orbital effects \cite{Alicea-Tunable}), and the non-Abelian character of the end Majorana modes can
 be probed by Josephson experiments.  Very recently it has been shown that the generic Luttinger Hamiltonian applicable to the
  \emph{hole-doped} nanowires also supports end Majorana modes \cite{zhang-tewari} in a manner similar to its electron-doped counterpart. The 1D electron- or hole-doped wires in the TS state can be arranged in a quasi-1D
network geometry \cite{Alicea} to test non-Abelian statistics \cite{Alicea} and perform  topological quantum computation (TQC) \cite{Sau-Universal,Beenakker} in the Bravyi-Kitaev (BK) scheme. \cite{Bravyi} In principle, TQC in the BK scheme is also possible with the 2D
 semiconductor TS states using Majorana fermion interferometry \cite{Interferometry} analogous to that in the $\nu=5/2$ fractional quantum Hall (FQH) states, \cite{Dassarma} chiral $p$-wave superconductors, \cite{Tewari-Strontium} and TS states on the surface of topological insulators. \cite{Fu,Akhmerov} Efforts to realize a semiconductor TS state with Majorana fermions in proximity to $s$-wave superconductors are currently
underway in many laboratories world-wide, concentrating on both 1D semiconducting nanowires and 2D semiconductor heterostructures
in close proximity to a regular bulk $s$-wave superconductor (e.g. Al, Nb).

In view of the ongoing experimental efforts on semiconductor TS states, it is important to understand the effects of disorder from the semiconductor surface on the realizability of the TS state.  (In recent works \cite{Stanescu} it has been established
that the disorder residing in the adjacent bulk superconductor has negligible effect on the semiconductor TS state, and therefore
in this paper we will ignore the effects of disorder from the bulk superconductor.) This is an especially important question \cite{potter,brouwer1,brouwer2} because the TS states in semiconductors explicitly break the time reversal (TR) symmetry and it is known that the superconducting quasiparticle gap in such systems, unlike that in TR-invariant superconductors, \cite{Anderson} is suppressed by disorder.
Below we will measure the extent of TR breaking in the semiconductor TS state (in both 1D and 2D) by the ratio $r=V_Z/(\alpha k_F)$, where $V_Z$ is the Zeeman energy in the TS state and $\alpha k_F$ gives the typical spin-orbit energy scale (with $\alpha$ as the Rashba spin-orbit coupling constant).
We will show that the disorder suppression of the superconducting quasiparticle gap from its clean value increases with increasing values of $r$.
Nonetheless, by giving explicit numbers we show that a reasonable TS state quasiparticle gap can be experimentally achievable in
 both electron- and hole-doped systems even in the presence of realistic disorder.
 For electron-doped wires we find that
 with a mobility of $\sim 100,000$ cm$^2$/V-s a measurable robust TS state gap of $50-100$ mK is achievable.
 We also find that for hole-doped wires with a
significantly larger spin-orbit energy $E_{SO}\sim 30$ meV,~\cite{winkler2,bernevig_zhang} a larger gap of $0.8$ K is achievable for a
mobility of $\sim 100$ cm$^2$/V-s.

In what follows, by $\Delta_s$ we denote the quasiparticle gap (pair potential) in the (proximity-inducing) bulk superconductor.
 For the semiconductor (2D or 1D), $\alpha$ is the Rashba spin-orbit coupling constant, $m^*$ is the effective mass (of electrons or holes),
 $V_Z$ is the Zeeman splitting required for the TS state, $\Delta_0$ is the proximity-induced pair potential,
 $\Delta$ is the quasiparticle gap at the Fermi surface, $k_F$ is the Fermi momentum ($\hbar = e = k_B =1$),
 and $1/\tau_{sm}$ is the effective intrinsic semiconductor disorder scattering rate.
 In the semiconductor, the combination of the Zeeman splitting and
the spin-orbit coupling results in a Fermi-surface quasiparticle gap $\Delta$, which differs from the proximity-induced pair
 potential $\Delta_0$, and thus we have three distinct superconducting gaps $\Delta_s$, $\Delta$ and $\Delta_0$ to
consider in this problem.

 In Sec.~II we first give our starting Hamiltonian for the semiconductor which includes a proximity-induced pair potential, an externally
  applied Zeeman splitting and a disorder potential.
 We then calculate the dependence of the disordered TS state quasiparticle gap on
 the semiconductor disorder scattering rate $1/\tau_{sm}$ (Eq.~\ref{scbornsol}).
In Sec.~III, we review the two systems that we will apply our
results to, namely, electron and hole-doped semiconductors.
 In section IV, we use the results of Sec.~II and Sec.~ III to give explicit realistic values for the various
 parameters for a stable TS state in electron-doped semiconductors.
 This is followed by the case of hole-doped semiconductors, where again using explicit values of the parameters we show that
 a stable TS state is possible. We find that the TS state gap in the hole-doped case can be much larger and more robust to
 disorder effects than in its electron-doped counterpart
 due to larger values of $\alpha$ and $m^*$.
 We conclude in Sec. V with a summary and an outlook with a detailed
qualitative discussion on why the key problem of choosing the optimal materials combination for realization of the TS
state is experimentally important.

\section{BdG Hamiltonian with disorder}
The BdG Hamiltonian in real-space for spin-$1/2$ electrons with
spin-singlet pairing potential is written as
\begin{equation}
H_{BdG}=\left(\begin{array}{cc}H_{sm}-\epsilon_{F,sm}&\Delta_0\\\Delta_0&\epsilon_{F,sm}-\sigma_y H_{sm}^*\sigma_y\end{array}\right)\label{BdGreal}
\end{equation}
where $\Delta_0$ is a real, constant, $s$-wave
 spin-singlet pairing potential in the semiconductor
which is proximity induced from an adjacent superconductor,
 $H_{sm}\equiv H_{sm}(\bm r,\sigma;\bm r'\sigma')$
is the non-interacting part of the semiconductor Hamiltonian with disorder  and $\epsilon_{F,sm}$ is the chemical
potential.
Here $\sigma_{x,y,z}$ are the Pauli matrices.
 For a disordered semiconductor, with spatially localised (i.e. short-ranged) spin-independent disorder,$H_{sm}$ can be written as
\begin{equation}
H_{sm}(\bm r,\sigma;\bm r'\sigma')=H_{sm,clean}(\bm r,\sigma;\bm r'\sigma')+V(\bm r)\delta(\bm r-\bm r')\delta_{\sigma\sigma'}\label{hsm}
\end{equation}
where $V(\bm r)$ is the local
potential induced by disorder.

In the limit of weak disorder and weak pairing where both $V(\bm r)$ and
$\Delta_0$ are smaller than the inter-band spacing of the spin-orbit bands in the semiconductor, it is convenient to
work in the basis of Bloch eigenstates $\ket{n,k}$ of $H_{sm,clean}(\bm r\sigma;\bm r'\sigma')$.
In particular, in the case where a single band, say $n$, crosses the Fermi-surface (which is necessary for realizing a topological superconducting state \cite{Zhang,Sato}), we can calculate
the low-energy spectrum of the Hamiltonian from an effective Hamiltonian which is obtained by taking
matrix elements of  the BdG Hamiltonian in Eq.~\ref{BdGreal} with respect to the Nambu spinors
\begin{equation}
\ket{n,\bm k}_{Nambu}=\left(\begin{array}{c}\ket{n,\bm k}\\\Theta \ket{n,-\bm k}\end{array}\right)
\end{equation}
where $\Theta=i\sigma_y K$  with $K$ being the complex conjugation
operator.
 In this approximation, the translationally invariant part
of the BdG Hamiltonian has the form
\begin{align}
H_0(\bm k,\bm k')=\left(\begin{array}{cc}\epsilon_{\bm k}-\epsilon_{F,sm}&\Delta_{\bm k}\\-\Delta^*_{-\bm k}&-(\epsilon_{-\bm k}-\epsilon_{F,sm})\end{array}\right)\delta(\bm k-\bm k')\label{h0}
\end{align}
where we have suppressed the band-index $n$, since we are restricted
to a single band and the term $H_{sm,clean}(\bm r\sigma;\bm r'\sigma')$
has been replaced by its eigenvalue $\epsilon_{\bm k}$ in the relevant
band. In the electron-doped case, $\epsilon_{\bm k}$ includes effects on
 the dispersion of the parameters $V_Z$ and $\alpha$ (see
Eq.~\ref{rashba}) while in the
hole-doped case, it additionally contains information about the Luttinger parameters (see Eq.~\ref{luttinger1}).
We note that, in spite of its apparent complexity involving the superconducting proximity
effect, spin-orbit coupling, Zeeman splitting, and disorder, we are still dealing here with an effective
exactly solvable one-particle quantum problem.

For weak superconducting pairing and disorder scattering, the TS quasiparticle gap is determined by
the energies and wave-functions near the Fermi energy $\epsilon_{F,sm}$.
 In particular, we will approximate the dispersion around the Fermi energy by
 $\epsilon_{\bm k}\approx \epsilon_{F,sm}+v_F(|\bm k|-k_F)$
where  $v_F$ is the Fermi-velocity.
Moreover, since the relevant states are near the Fermi wave-vector (i.e $|\bm k|\approx k_F$),
we will assume that matrix elements such as  $\expect{\bm k|\bm k_1}$ depend only on the
directions of the momenta $\bm k$ and $\bm k_1$.
To calculate the matrix elements of $\ket{\bm k}$, we will assume that the Bloch eigenstates have
a simple decomposition as
\begin{equation}
\expect{\bm r;\sigma|\bm k}\equiv u_{\bm k}(\bm r;\sigma)=e^{i\bm k\cdot\bm r}u_{\bm k}(\sigma).\label{bloch}
\end{equation}
 This assumption
is valid as long as all perturbations involve momentum transfers that are smaller than a Bloch vector.
For electrons $\sigma$ is simply the electron-spin.
When applied to holes $\sigma$ would represent the 4-component
pseudo-spin degree of freedom associated with the Luttinger model. \cite{zhang-tewari}

In this paper we will consider  both 1D and 2D systems, and these cases will turn out
to be closely related. In the  two-dimensional case, we will restrict ourselves to rotationally
symmetric systems with  circularly symmetric Fermi surfaces where $\bm k=(k_F\cos{\theta_{\bm k}},k_F\sin{\theta_{\bm k}})$.
Using the azimuthal symmetry of the Bloch Hamiltonian corresponding to $H_{sm,clean}$,
we can define a Hermitian matrix $R_z$, which generates $z$-axis rotations so that
\begin{equation}
\ket{|k|,\theta_{\bm k}}=e^{i (R_z-\lambda)(\theta_{\bm k})}\ket{|k|,0}\label{ket2D}
\end{equation}
where $\lambda$ is chosen so that $\ket{|k|,\theta_{\bm k}}$ is a
single valued function of $\theta_{\bm k}$.
Since Kramer's theorem requires $R_z$ to have half-integer eigenvalues $\lambda=1/2$.
For  one-dimensional systems, such as electron- or hole-doped semiconducting nanowires, we will consider systems  with inversion symmetric
Fermi points $\pm k_F$. In this case, the relevant states at the Fermi-level are $\ket{\pm k_F}$.
The 2D systems with Rashba spin-orbit coupling will turn out to have results that are solely dependent on the
wave-functions only at $\ket{k_F,\theta_{\bm k}=0,\pi}$, which are identical in form to the states $\ket{\pm k_F}$
relevant for the 1D case. Therefore, these two cases are closely related and for results such as the quasiparticle gap in
 Eqs.~\ref{eq:x} and ~\ref{scbornsol},
we will only state the result for the 1D case and imply a similar result in 2D, which can be derived easily.

The superconducting pairing potential $\Delta_{\bm k}$  in
the projected Hamiltonian in  Eq.~\ref{h0} is the matrix element of $\Delta_0$ in Eq.~\ref{BdGreal} between states near the
Fermi-level, which is written as
\begin{equation}
\Delta_{\bm k}=\Delta_0\bra{ \bm k}\Theta\ket{-\bm k}.\label{deltak}
\end{equation}
Using Eq.~\ref{ket2D} in the rotationally symmetric 2D case, $\Delta_{\bm k}$ simplifies to
\begin{align}
&\Delta_{\bm k}=\Delta_0\bra{ \bm k}\Theta\ket{-\bm k}=\Delta_0\bra{0}e^{-i(R_z-\lambda)\theta_{\bm k}}\Theta e^{i(R_z-\lambda)\theta_{\bm k}}\ket{\pi}\nonumber\\
&=e^{2 i\lambda\theta_{\bm k}}\Delta_0\bra{0}\Theta\ket{\pi}=\Delta_0\bra{0}\Theta\ket{\pi} e^{i\theta_{\bm k}}\label{deltak1},
\end{align}
where $\theta_{\bm k}$ is the angle of the wave-vector $\bm k$ in the $k_x-k_y$ plane.
Here we have assumed that $R_z$, which transforms $k_F\rightarrow -k_F$, is odd under time-reversal symmetry.
The TS quasiparticle gap for the Hamiltonian Eq.~\ref{h0}, in the absence of disorder (i.e. $V(\bm r)=0$),
is given by
\begin{equation}
\Delta=|\Delta_{\bm k}|.\label{delta}
\end{equation}

The disorder scattering, induced by the potential $V(\bm r)$ in Eq.~\ref{hsm}, can be projected
into states near the Fermi energy in a similar way as
\begin{align}
V(\bm k,\bm k')=\left(\begin{array}{cc}v(\bm k,\bm k')&0\\0&-v^*(-\bm k,-\bm k')\end{array}\right)\label{Vmat}
\end{align}
where $v(\bm k,\bm k')=\bra{\bm k}V(\bm r)\ket{\bm k'}$.
For spatially uncorrelated white noise disorder (i.e. $\expect{V(\bm r)V(\bm r')}=v^2\delta(\bm r-\bm r')$),
and using the simple form for the Bloch functions Eq.~\ref{bloch},
the disorder propagator $\expect{v(\bm k_1,\bm k_2)v(\bm v_3,\bm k_4)}$ can
be written as
\begin{equation}
\expect{v(\bm k_1,\bm k_2)v(\bm v_3,\bm k_4)}=v^2\expect{\bm k_1|\bm k_2}\expect{\bm k_3|\bm k_4}.\label{disordervertex}
\end{equation}
The disorder strength $v^2$ is related to the mean-scattering time (i.e. momentum relaxation time)
\begin{equation}
\tau_{sm}=\frac{1}{\pi v^2 N(0)},\label{tausm}
\end{equation}
which in turn can be determined experimentally from
mobility measurements using the equation
\begin{equation}
\mu_{sm} \approx \frac{e\tau_{sm}}{m^*},\label{musm}
\end{equation}
where $m^*$ is the transport effective mass and $N(0)$ is the density of states at the Fermi
 surface. Note that we have dropped a matrix-element factor in determining the scattering time in
Eq.~\ref{tausm}. This is partly to account for multi-band effects that we are not directly accounting
for in this paper. Since we are using mobility only as
a qualitative estimator of the scattering, the combination of Eqs.~\ref{tausm},~\ref{musm} provide a good
estimate of the disorder scattering $v^2$. Although Eqs. 12 and 13 strictly apply only in the case of
white noise disorder arising from short-ranged impurity scattering, we assume that they remain valid for long-range
disorder as well in the presence of random charged impurity scattering since carrier screening of the
impurity potential should render the Coulomb disorder into an
effective short-range disorder. In addition, our use of $\tau_{sm}$ to operationally
characterize the semiconductor disorder should be qualitatively valid for any kind of
impurity potential.  While we have made a single Fermi surface approximation in Eq.~\ref{h0},
this is more for the sake of simplicity. In the next paragraph,
 we will comment on how our results generalize to the multiple Fermi
surface case, as for example in the case of multi-band occupancy in the semiconductor
nanowire where the Fermi level could lie in some high sub-band rather than the ground 1D
sub-band. \cite{Stanescu}

The topological superconducting quasiparticle gap in the presence of a disorder potential may be
determined by calculating the disorder averaged Green function within the Born approximation.\cite{agd}
The disorder averaged Green function can be calculated from the self-energy $\Sigma$
using the Dyson equation
\begin{equation}
G^{-1}(\bm k,\omega)=G_0^{-1}(\bm k,\omega)-\Sigma(\bm k,\omega),\label{dyson}
\end{equation}
where the Green function for the clean system is given by
\begin{equation}
G_0^{-1}(\bm k,\omega)=(\omega-H_0(\bm k))=\omega-(\epsilon_{\bm k}-\epsilon_{F,sm})\tau_z-\Delta_{\bm k}\tau_+-\Delta_{\bm k}^*\tau_-.\label{g0}
\end{equation}
The self-energy $\Sigma(\bm k,\omega)$ is approximated within the self-consistent Born approximation\cite{agd} as
\begin{equation}
\Sigma_{\alpha\beta}(\bm k,\bm k')=\int d\bm k_1 \expect{V_{\alpha\lambda}(\bm k,\bm k_1)
V_{\delta\beta}(\bm k_1,\bm k)}G_{\lambda\delta}(\bm k_1)\delta(\bm k-\bm k')\label{sigma}
\end{equation}
where $\expect{V_{\alpha\lambda}(\bm k,\bm k_1)V_{\delta\beta}(\bm k_1,\bm k)}$ is the disorder propagator from
 Eqs.~\ref{Vmat},~\ref{disordervertex},  $\alpha,\beta,\lambda,\delta$ are Nambu indices,
 and for brevity of notation we have dropped
the $\omega$ dependence of $\Sigma$ and $G$ in this equation.
The self-energy within the first-order Born approximation is obtained by replacing $G$ by $G_0$ in Eq.~\ref{sigma}.
The disorder calculation for the case of multiple Fermi-surfaces follows in an analogous way. Since momentum is a good quantum
number in Eq.~\ref{dyson}, Eq.~\ref{dyson} does not couple the multiple Fermi-surfaces and therefore can be solved
for each Fermi surface separately. The self-energy term Eq.~\ref{sigma} in principle has contributions from all Fermi surfaces.
However, it is dominated by the contribution from the Fermi-surface with the smallest gap. Therefore, our calculation, which
focuses on the Fermi surface with the smallest gap, is expected to yield qualitatively correct results.

Solving the Dyson equations in the first-order Born approximation in the 1D and 2D cases, as discussed in appendix \ref{firstorderborn},
we obtain the disorder averaged TS quasiparticle gap
\begin{equation}
E_g=\Delta\left[1-\frac{1}{2^{1/3}}\left(\pi\tau_{sm}\Delta\right)^{-2/3}|\expect{k_F|-k_F}|^{4/3}\right].\label{eq:x}
\end{equation}
Note that the disorder renormalization of the gap vanishes in the
time-reversal symmetric limit since in that case $\ket{\bm k}$ and
$\ket{-\bm k}$ are Kramer's pairs and $\expect{k_F|-k_F}=0$.
The result Eq.~\ref{eq:x} gives the reduction in the quasiparticle gap due to weak disorder
 within the first-order Born approximation.

 The Dyson equations, Eqs.~\ref{dyson},~\ref{sigma}, can in fact be solved for arbitrary disorder strengths within
the self-consistent Born approximation as explained in Appendix.~\ref{selfconsistentborn}. The resulting
TS state quasiparticle gap in the presence of disorder is
found to be
\begin{align}
E_g=\Delta\sqrt{1-3 x+3 x^2-3 x^3}\label{scbornsol}
\end{align}
where $x=\left(\pi\tau_{sm}\Delta\right)^{-2/3}|\expect{k_F|-k_F}|^{4/3}$.
Note that the gap $E_g$ within the first-order Born approximation in Eq.~\ref{eq:x} is not simply obtained by taking the limit of 
small $x$ in the self-consistent Born result Eq.~\ref{scbornsol}. This is because the Born approximation is a series 
expansion in terms of the Green function, which has a pole at $\omega\sim E_g$, so that the expansion parameter in 
the series expansion diverges (at any value of $x$) at the relevant point and one cannot expect the lowest order 
approximation to match the self-consistent result. Despite, this technical point, the first-order and self-consistent 
Born approximation results, Eq.~\ref{eq:x} and Eq.~\ref{scbornsol} are qualitatively similar at small values of the 
parameter $x$.  

The above equation is valid as long as the expression under the square root is positive. In fact, the quasiparticle
gap $E_g$ closes for sufficiently strong disorder (or sufficiently small $\Delta\tau_{sm}$) so that, by solving for $E_g=0$ in Eq.~\ref{scbornsol}, the disorder scattering time is found to satisfy the lower bound
\begin{equation}
\tau_{sm}>\frac{3.4}{\pi}\hbar \Delta^{-1} |\expect{k_F|-k_F}|^{2}\label{taut},
\end{equation}
above which the TS state quasiparticle gap $E_g>0$.
In this paper, we will also assume that $\tau_{sm}>1/\tilde{\epsilon}_{F,sm}$ ($\tilde{\epsilon}_{F,sm}$ is the Fermi
 energy with respect to the bottom of the top-most filled band) need not be strictly enforced because of the
existence of multiple bands in the actual experimental system, which will evade the localization problem.

From Eqs.~\ref{musm} and~\ref{taut}, we find that
 the TS state gap induced on the Fermi surface survives as long as the mobility
of the system exceeds the threshold
\begin{equation}
\mu_{sm}>3.4 \frac{e \hbar}{\pi m^*\Delta} |\expect{k_F|-k_F}|^{2}.\label{mut1}
\end{equation}
This suggests that the mobility threshold can be decreased by
decreasing the Kramer's pair overlap $|\expect{k_F|-k_F}|$,  which, as we show in Eq.~\ref{krameroverlap}, can be achieved by
reducing the Zeeman potential. We will apply all these results to electron and hole-doped semiconductors
in the following two sections.

\section{Proximity-coupled electron and hole-doped semiconductors}
In this section we review the two recently proposed systems, electron- and hole-doped semiconductors, to which we will apply
our results derived in the previous section.
\subsection{Electron-doped semiconductors}
For 2D electron-doped semiconductors with Rashba and Zeeman couplings, $H_{sm,clean}$ in Eq.~\ref{hsm}
can be written as
\begin{equation}
H_{sm,clean}(k)=\frac{k^2}{2 m^*}+\alpha (\bm k\times \bm\sigma)\cdot \hat {\bm z}+V_Z \sigma_y-\epsilon_{F,sm}\label{rashba}
\end{equation}
where $m^*$ is the effective mass of the electrons, $\alpha$ is the Rashba spin-orbit coupling constant, and $V_Z$ is the
Zeeman splitting. For a one-dimensional nanowire along the $y$-axis, the above Hamiltonian reduces to
\begin{equation}
H_{sm,clean}(k)=\frac{k^2}{2 m^*}+\alpha k\sigma_x+V_Z \sigma_y-\epsilon_{F,sm}.\label{rashba1D}
\end{equation}

For a chemical potential $|\epsilon_{F,sm}|<V_Z$, the above Rashba spin-orbit-coupled Hamiltonian
has a single band at the Fermi-level with an eigenstate described by the   spinor
\begin{align}
&\ket{k_F}=\frac{1}{\sqrt{2 [V_Z^2+\alpha^2 k_F^2+V_Z\sqrt{V_Z^2+\alpha^2k_F^2}]}}\nonumber\\
&\left(\begin{array}{c}\alpha k_F\\-\sqrt{V_Z^2+\alpha^2k_F^2}-V_Z\end{array}\right).
\end{align}
Now, using Eq.~\ref{delta}, the superconducting
quasiparticle gap in the clean electron-doped case is given by
\begin{equation}
\Delta\sim\frac{\alpha k_F}{\sqrt{V_Z^2+\alpha^2k_F^2}}\Delta_0.\label{delta1}
\end{equation}
The proximity-induced pairing potential in the
semiconductor, $\Delta_0$, can be related to the pairing potential
in the bulk superconductor $\Delta_s$ by the relation \cite{robustness}
\begin{equation}
 \Delta_0=\Delta_s\frac{\lambda}{\lambda+\Delta_s}\label{delta0}
\end{equation}
where
\begin{equation}
\lambda=\pi |t|^2\rho_{SC}(E_F).\label{lambda}
\end{equation}
 Here $t$ represents
the tunneling matrix element between the semiconductor and the superconductor, $\rho_{SC}(E_F)$
 is the normal state density of states of the
superconductor and $\lambda$ gives the rate at which electrons from the
semiconductor tunnel into the superconductor. \cite{robustness}
We note that for weak-tunneling $\Delta_0\sim \lambda$, and for strong tunneling, $\Delta_0\sim\Delta_s$. 
The superconducting proximity effect in electron-doped InAs/superconducting Al interfaces has been already been 
observed to lead to $\Delta_0\sim \Delta_s\sim 2$ K, ~\cite{electron_proximity} so that one can infer that the transparency 
parameter $\lambda\gtrsim \Delta_s$.
The Kramer-pair overlap $\expect{k_F|-k_F}$, which determines the effect of disorder scattering via Eq.~\ref{scbornsol}, is
then given by
\begin{equation}
\expect{k_F|-k_F}=\frac{V_Z}{\sqrt{V_Z^2+\alpha^2k_F^2}}\label{krameroverlap}
\end{equation}
which in the limit $\frac{V_Z}{\alpha k_F}\ll 1$ becomes $\expect{k_F|-k_F}\sim \frac{V_Z}{\alpha k_F}$.

\subsection{Hole-doped semiconductors}
Recently, it has also been proposed~\cite{zhang-tewari} that hole-doped
nanowires, because of the possibly stronger spin-orbit coupling~\cite{winkler2,bernevig_zhang} and larger effective mass of holes,
might be a better candidate for creating topological superconductors and Majorana fermions. 
The hole-bands of a group III-V semiconductor such as InAs or GaAs are composed of $p$-orbitals, which have 
orbital angular momentum $L=1$. Because of strong spin-orbit coupling and cubic symmetry of these semiconductors, the 
description of the states near the top of the valence band at $\bm k=0$ must be written in terms of spin-$3/2$ total 
angular momentum matrices $\bm J=\bm L+\bm S$. The total spin-1/2 manifold of states forms a split-off band, which is 
separated by a large energy and therefore can be ignored.
 The resulting three-dimensional Hamiltonian is known as the four-band Luttinger model~\cite{luttinger},
 which is written as 
\begin{align}
&H_{3D}=\frac{1}{2 m_0}\left[\left(\lambda_1+\frac{5}{4}\lambda_2\right)p^2-2\lambda_2\sum_{\alpha=x,y,z}p_\alpha^2J_\alpha^2\right]\nonumber\\
&-\lambda_3\frac{1}{2 m_0}\sum_{\alpha\neq\beta=x,y,z}p_\alpha p_\beta J_\alpha J_\beta,
\end{align} 
where $\bm p=-i\bm \nabla$ and $\gamma_{1,2,3}$ are effective Luttinger parameters.
The effective mass $m_0$ for the holes is negative so that the curvature of the valence bands are inverted with
 respect to the conduction bands.
We will consider a nanowire formed by confining the bulk hole-doped semiconductor confined strongly along the $x$ and
 $z$-directions, 
so that the confinement energy scale $\sim \frac{1}{2 m_0 L_y^2}$ is large compared to all energy-scales we will be interested in. 
Therefore one can project the three-dimensional Hamiltonian $H_{3D}$ into the lowest confinement band of the nanowire by replacing 
$p_{x,z}\rightarrow \expect{p_{x,z}}=0$, $p_{x,z}^2\rightarrow \expect{p_{x,z}^2}=\frac{\pi^2}{L_z^2}$ and adding
 a Rashba spin-orbit coupling 
proportional to $\alpha$~\cite{rashba,winkler1,winkler2} and a Zeeman term proportional to $V_Z$ to account for electric field
 induced inversion symmetry breaking and 
magnetic field-induced time-reversal symmetry breaking respectively.    
The resulting Hamiltonian for a hole-doped
nanowire along $y$ with a Rashba coupling 
 (for the lengths in the transverse directions $L_z=L_x$) is given by, \cite{zhang-tewari}
\begin{align}
&H_{sm,clean}(k)=-(\frac{\lambda_1}{2}+\frac{5\lambda_2}{4}-\lambda_2 J_y^2)k^2-\alpha k J_x\nonumber\\
&+\frac{2\beta_2}{\sqrt{3}}(\frac{15}{4}-J_y^2)+V_Z J_y-\epsilon_{F,sm}\label{luttinger1}
\end{align}
where $\beta_2=\frac{\pi^2\lambda_2\sqrt{3}}{2 L_z^2}$.
The Hamiltonian at $k=0$, commutes with $J_y^2$. Moreover, the states with $J_y^2=1/4$ 
form the highest energy states in the valence band. These states also have the lower effective mass of the
 two manifolds $J_y^2=1/4,9/4$ 
of states. Note that this is opposite to the two-dimensional case, where the heavy-hole is closer to the 
band-edge. Thus the light-hole states with $J_y=\pm 1/2$ form the set of states closest to the band-edge and 
 we can restrict to this two-dimensional subspace $J_y=\pm 1/2$, where we identify $J_y=\sigma_y=\pm 1/2$ as
 a pseudo-spin variable.
 In this subspace, $J_x\sim \sigma_x$
and $J_z\sim \sigma_z$, so that the Hamiltonian for the
two upper bands effectively becomes
\begin{equation}
H_{sm,clean}(k)=-(\frac{\lambda_1}{2}+\lambda_2)k^2-\alpha k \sigma_x+V_Z \sigma_y+\epsilon_{F,sm}.\label{rashbahole}
\end{equation}

The hole Hamiltonian Eq.~\ref{rashbahole}, in the specific isotropic limit we consider,
 is identical (up to sign) to the one-dimensional electron-doped Rashba Hamiltonian in Eq.~\ref{rashba1D} and 
therefore one can expect most of the physics discussed in the previous sub-section 
for the electron-doped nano-wire in this limit to translate to the hole-doped case, except for the different effective 
mass and Rashba spin-orbit coupling. While the band-structure of the hole-doped wire is quite similar to the electron-doped 
case, it is not obvious from this discussion that the $s$-wave proximity effect has the same form in the $\sigma$ spin-space.
In principle, the proximity-induced effective superconducting pairing potential $\Delta_0$ at an interface between the
 semiconductor and the superconductor can be calculated 
 by projecting the anomalous self-energy $\Sigma_{SC}$ induced by proximity to the superconductor,
 ~\cite{robustness} into the low-energy space 
of states $\ket{\sigma_y=\pm 1/2}$. Since these states form a Kramers pair, one can see that time-reversal
 symmetry only allows a conventional 
spin-singlet pairing in the $\sigma_y=\pm 1/2$ space.  
Furthermore, the sub-space of states in the $\ket{\sigma_y=\pm 1/2}$ basis may be expanded in terms
 of the original spin eigenstates
 $\ket{m_S=\pm \frac{1}{2}}$ and
$\ket{L_{x,y,z}=0}$ (which are related to the orbital angular momentum eigenstates 
 $\ket{m_L=0,\pm 1}$
 by a unitary transformation)
 as  
\begin{align}
&\ket{\sigma_y=+\frac{1}{2}}\equiv \ket{J_y=\frac{1}{2}}\nonumber\\
&=\frac{1}{\sqrt{3}}[\ket{L_x=0,S_y=-\frac{1}{2}}+i\ket{L_z=0,S_y=-\frac{1}{2}}\nonumber\\
&-\ket{L_y=0,S_y=\frac{1}{2}}]\\
&\ket{\sigma_y=-\frac{1}{2}}\equiv \ket{J_y=-\frac{1}{2}}\nonumber\\
&=\frac{1}{\sqrt{3}}[\ket{L_x=0,S_y=\frac{1}{2}}-i\ket{L_z=0,S_y=\frac{1}{2}}\nonumber\\
&-\ket{L_y=0,S_y=-\frac{1}{2}}].
\end{align}
To calculate $\Delta_0$, which is related to the matrix-element $\bra{\sigma_y=+1}\Sigma_{SC}\ket{\sigma_y=-1}$
 of a spatially smoothly varying (on the orbital scale)
 proximity-induced self-energy $\Sigma_{SC}$, 
one can ignore the matrix elements involving the orbital $\ket{L_x=0},\ket{L_z=0}$ for SC interfaces normal to 
the $\hat{z}$-axis.
 This leads to the effective barrier transparency parameter $\lambda$ 
in Eq.~\ref{delta0} to be reduced by a factor of $3$. 
The corresponding pairing potential $\Delta_0$ from Eq.~\ref{delta0} is not significantly affected and remains the same between electrons 
and holes provided a moderately transparent barrier with $\lambda>\Delta_s$ can be achieved.
 Therefore, more generally, one can expect a proximity-effect where the transparency-factor
 of the Sm/SC interface, $\lambda$, is suppressed by at most a factor of order 1, which
a value for the pairing potential $\Delta_0\sim \Delta_s$, even for holes.

To summarize, while the hole-doped nanowires have fundamentally different microscopic Hamiltonians consisting of a triplet of 
orbitals with $p$-symmetry, for nanowires it is possible to work in the limit of symmetric confinement along the two directions transverse to the wire, where we have shown that the 
effective Hamiltonian and proximity-effect are identical in form to the electron-doped wire
 case. Of course, the parameters in the Hamiltonian 
are expected to be quite different and in particular the effective mass and Rashba spin-orbit coupling are expected to be much larger. ~\cite{winkler2,bernevig_zhang} At the same time, the corresponding mobility for the semiconductors that can be achieved are found to be significantly lower, ~\cite{holemobility} 
 indicating much stronger disorder for hole-doped systems.  Thus, the hole systems have the advantage (disadvantage) of
 stronger SO coupling (larger disorder), and whether the net prospect for realizing the non-Abelian Majorana mode is greater
 in hole-based systems (than the electron-based systems) or not will depend entirely on which of these two effects wins over
 in realistic samples.  Another possible advantage of the hole systems is the higher value of the Lande g-factor which
 necessitates weaker magnetic field values in achieving the condition for topological superconductivity. One should,
 however, realize that very little actual experimental work exists in hole-based semiconductor nanowires.
However,
it is encouraging that the superconducting proximity effects in both electron and hole-based semiconductors have been observed 
in experiments.~\cite{electron_proximity,hole_proximity}
  Therefore,  we do believe, 
 that the hole-based systems (e.g. InAs, InSb, SiGe, and perhaps even GaAs) should be included in the experimental search
 for the Majorana mode in semiconductor-superconductor heterostructures.

\section{TS state quasiparticle gap in the presence of disorder}
The  clean-limit quasiparticle gap $\Delta$ in the topological superconducting state can be obtained for the Rashba spin-orbit coupled Hamiltonian
Eq.~\ref{rashba} using Eqs.~\ref{delta1} and ~\ref{delta0} as
\begin{equation}
\Delta=\frac{\lambda \Delta_s}{\lambda + \Delta_s} \frac{\alpha k_F}{\sqrt{V_Z^2+\alpha^2k_F^2}}.\label{delta5}
\end{equation}
This expression suggests that to obtain a large gap, $\Delta$, in the semiconductor one should
restrict oneself to the limit where $\alpha k_F\gtrsim V_Z$.
 Moreover, to remain in the requisite odd sub-band (confinement bands) limit it is also necessary to choose  a
chemical potential $\epsilon_{F,sm}$ in the gap (at $\bm k=0$) induced by the
Zeeman splitting $V_Z$.\cite{long-PRB} This restricts $k_F\approx \frac{m^* \alpha}{2}$ in the highest filled
sub-band. In this
limit one can approximate $\alpha k_F\approx E_{SO}=2 m^*\alpha^2$,  which we call the spin-orbit energy $E_{SO}$.
This spin-orbit energy $E_{SO}$ is a crucial energy scale in the problem which is to be compared with the other independent energy scales such as
$V_Z$, $\Delta_{s}$ and $1/\tau_{sm}$.
 Therefore, from Eq.~30, to obtain a large quasiparticle
gap at the Fermi surface the system must be restricted to the regime
\begin{equation}
V_Z\lesssim E_{S0}.\label{vzcons}
\end{equation}

The semiconductor system is in the topological state if and only if the Pfaffian invariant of the relevant BdG Hamiltonian
is negative, \cite{parag}
which is equivalent to the topological condition
\begin{equation}
\tilde{V}_Z^2>\tilde{\Delta}_0^2+\tilde{\epsilon}_{F,sm}^2\label{topcons1}
\end{equation}
where $\tilde{V}_Z=V_Z/(1+\lambda/\Delta_s)$, $\tilde{\Delta}_0=\Delta_0$ and $\tilde{\epsilon}_{F,sm}=\epsilon_{F,sm}/(1+\lambda/\Delta_s)$
 are the renormalized
parameters in the effective Hamiltonian of the proximity coupled semiconductor after the superconductor has
been integrated out. \cite{robustness,long-PRB} Thus, using Eq.~\ref{delta0} and Eq.~\ref{topcons1}, the topological condition in terms of the bare parameters can be written as,
\begin{equation}
V_Z^2>\lambda^2+\epsilon_{F,sm}^2.\label{topcons}
\end{equation}
Eq.~\ref{topcons} and Eq.~\ref{vzcons}, together with Eq.~\ref{scbornsol} (or Eq.~\ref{mut1}), define the optimal constraints on
the material considerations for the realization of the TS phase in the
semiconductor.

In what follows, we assume that the system is reasonably deep in the topological superconducting phase
 (i.e. $V_Z>2\lambda,2 \epsilon_{F,sm}$) so the smallest gap occurs near the
 Fermi wave-vector $|\bm k|\sim k_F$ instead of near $|\bm k|\sim 0$.\cite{long-PRB}
Therefore in attempting to optimize the quasiparticle gap one notices
 from Eqs.~\ref{vzcons}, \ref{topcons} a hierarchy of energy scales
\begin{equation}
E_{SO}> V_Z > 2 \lambda>2 \Delta> 2 E_g\label{ineq}.
\end{equation}
The first two-inequalities from the left are constraints between the external parameters $E_{SO},V_Z$ and $\lambda$, out of 
which $V_Z$ and $\lambda$ must be tuned to satisfy the above inequality. The other external parameter is the disorder 
scattering rate in the semiconductor, $\tau_{sm}^{-1}$, which  must be reduced to satisfy $\Delta> (\pi\tau_{sm})^{-1}\sqrt{1+\frac{E_{SO}^2}{V_Z^2}}$.
The two-inequalities on the right-hand side of Eq.~\ref{ineq} are between the derived quantities 
$\Delta$ and $E_g$, which are completely determined once $E_{SO},V_Z,\lambda,\tau_{sm}$ are given, 
so that  the condition $2\lambda > 2\Delta$ and $2\Delta>2 E_g$ follow as direct consequences of
 Eqs.~\ref{delta5} and Eq.~\ref{scbornsol}, respectively.
Therefore, the disordered TS state quasiparticle gap, $E_g$, is
constrained by the spin-orbit energy scale $E_{SO}$, which must be as large as possible in the problem.

In the presence of disorder, one can use
 Eq.~\ref{scbornsol} together with Eq.~\ref{krameroverlap} to show
  that for $V_Z\lesssim E_{SO}$,
 the disorder renormalized TS state quasiparticle
gap is given by
\begin{equation}
E_g=\Delta\sqrt{1-3 x+3 x^2-3 x^3}\label{qpgap}
\end{equation}
where $x=\left(\pi\tau_{sm}\Delta\right)^{-2/3}\left(\frac{V_Z}{\sqrt{V_Z^2+E_{SO}^2}}\right)^{4/3}$.
The mobility threshold, which is the minimum mobility required to realize a
non-zero TS state gap $E_g$ in the presence of disorder, is obtained by using Eqs.~\ref{mut1},~\ref{krameroverlap}
 to be
\begin{equation}
\mu_{sm}>6.8 \frac{e \hbar  V_Z (\Delta_s+V_Z/2)}{\pi m^* \Delta_s E_{SO} \sqrt{V_Z^2+E_{SO}^2}},\label{mut2}
\end{equation}
which vanishes in the limit $V_Z\rightarrow 0$. Note we have reintroduced the $\hbar=1$ to make the expression 
dimensionally consistent. Since $\mu_{sm}$ increases with $\Delta$, which in turn increases with $\lambda$, for deriving 
Eq.~\ref{mut2}, we have 
assumed the $\lambda$ is chosen to reach its maximum value $\lambda=V_Z/2$, above which the system can no longer be considered 
to be deep in the topological phase (see Eq.~\ref{ineq}). In fact, we will assume that  $V_Z$ has been tuned to be $V_Z=2 \lambda$ 
for the rest of the paper.

\begin{figure}
\centering
\includegraphics[scale=0.4,angle=270]{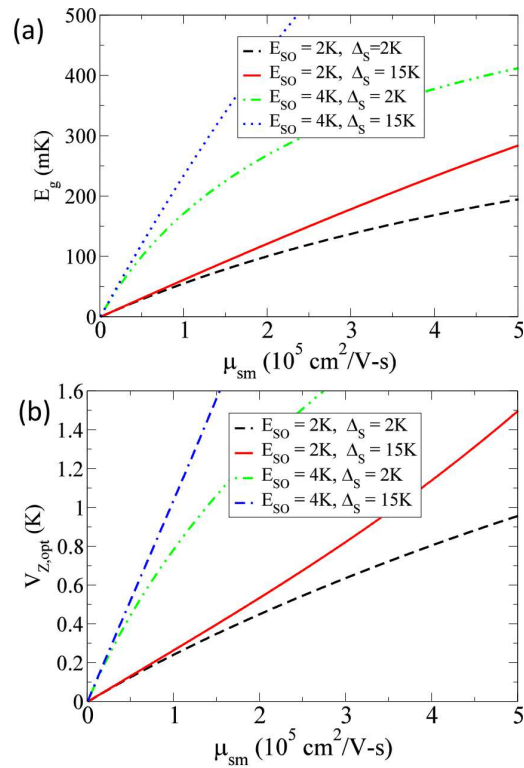}
\caption{(a) Calculated disordered TS state quasiparticle gap $E_g$ as a function of semiconductor mobility  $\mu_{sm}$ for
 electron-doped wires with $E_{SO}=2 m^*\alpha^2 = 2K$, \cite{kouwenhovenprivate} and $E_{SO}=4 K$ allowing for 
future electric-field enhanced spin-orbit coupling.~\cite{Zhang-Xia} (b) The values of $V_{Z}=2 \lambda$ realizing $E_g$ in panel (a) plotted with $\mu_{sm}$.
The gaps in the adjacent bulk superconductor are taken to be $\Delta_s=2 K$ corresponding to Al and $\Delta_s=4 K$
corresponding to Nb.}\label{Fig1}
\end{figure}

\begin{figure}
\centering
\includegraphics[scale=0.4,angle=270]{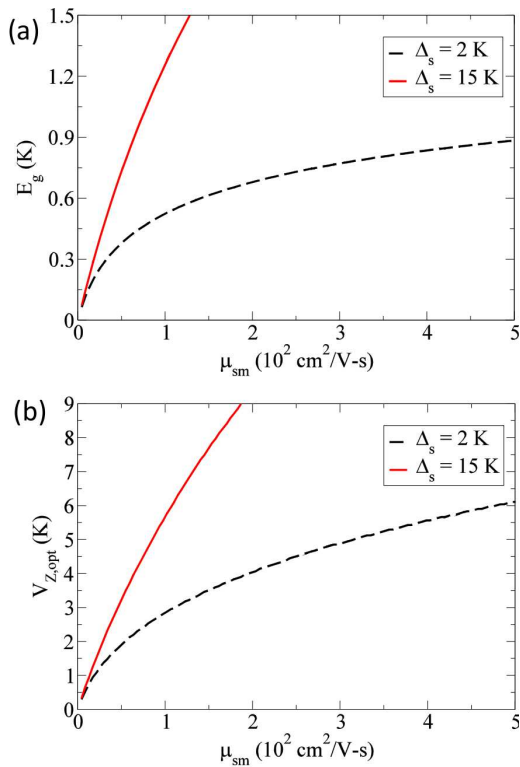}
\caption{(a) Calculated disordered TS state quasiparticle gap $E_g$ as a function of semiconductor mobility  $\mu_{sm}$ for
 hole-doped wires with $E_{SO}= 300 K$.~\cite{winkler2,bernevig_zhang} (b) The values of $V_{Z}=2 \lambda$ realizing $E_g$ in panel (a) plotted with $\mu_{sm}$.
The gaps in the adjacent bulk superconductor are taken to be $\Delta_s=2 K$ corresponding to Al and $\Delta_s=4 K$
corresponding to Nb.}\label{Fig2}
\end{figure}

 Of course, from Eq.~\ref{topcons},
there is a threshold value for $V_Z$ only above which the TS state itself is realized.
Therefore, $V_Z$ cannot be taken as zero in Eq.~\ref{qpgap} to maximize the disordered TS state gap $E_g$. In Eq.~\ref{qpgap}, to deduce
 the optimum $E_g$, we use the r.h.s of Eq.~\ref{delta5} to substitute for $\Delta$ and, for a robust topological state, we take $\lambda=V_Z/2$. The r.h.s of Eq.~\ref{qpgap} is then a function of $V_Z$, $\tau_{sm}$ (or $\mu_{sm} \sim e\tau_{sm}/m^*$),
 $\Delta_s$ and $E_{SO}$. Among these $\Delta_s$ (the gap in the adjacent superconductor) and $E_{SO}$ (the spin-orbit
coupling energy in the semiconductor) are known experimental quantities. Therefore  Eq.~\ref{qpgap} defines a
formula for $E_g$ in terms of $V_Z$ parametrically related by $\mu_{sm}$. For given values of $\mu_{sm}$, we maximize $E_g$
with respect to $V_Z$ and plot these maximized values of $E_g$ as a function of $\mu_{sm}$ for electron and hole-doped
 semiconductors in Figs.~\ref{Fig1} and \ref{Fig2}, respectively. These plots are the central results of this paper.

From Fig.~\ref{Fig1} it is clear that for electron-doped wires
 with a mobility of $100,000$ cm$^2$/V-s \cite{electronmobility},
which has been achieved for wires with a diameter $60$ nm, a measurable robust TS state gap of $50-100$ mK is achievable.
As is also clear from Fig.~\ref{Fig2}, for hole-doped wires with a
significantly larger spin-orbit energy $E_{SO}\sim 30$ meV, a larger gap of $0.8$ K is achievable for a
mobility of $100$ cm$^2$/V-s. \cite{holemobility}
It should be noted that in general the holes have much lower mobilities than electrons in the same semiconductor
 material even after taking into account their large effective mass difference.

\section{Summary and Conclusion}
In this paper we have addressed the effects of disorder from the semiconductor surface on the stability of the topological superconducting state on 1D and 2D electron- and hole-doped
semiconductors proximity coupled to bulk 3D $s$-wave superconductors. In recent works \cite{Stanescu} it has been shown that the effects of disorder from the adjacent bulk superconductor on the TS state is minimal, so we have ignored this effect. However, since the TS
state in the semiconductor explicitly breaks the time-reversal symmetry (due to an external Zeeman splitting), even the disorder from the
semiconductor surface itself has an effect on the stability of the TS state and will close the BCS-like superconducting quasiparticle gap at the
semiconductor Fermi surface for low enough mobility.
We have shown that the disordered TS state quasiparticle gap $E_g$ is suppressed with increasing values of the ratio $r$ between the Zeeman splitting $V_Z$ and the spin-orbit energy scale $E_{SO}=2 m^* \alpha^2$, $r=V_Z/E_{SO}$. The dimensionless quantity $r$ measures the
extent of the TR symmetry breaking in the semiconductor to produce the TS state and Majorana fermions.

Our main results are plotted in Figs.~\ref{Fig1} and ~\ref{Fig2}.
 They display the achievable disorder renormalized TS state gap $E_g$ as a function of the
  semiconductor mobility $\mu_{sm}$ in both electron and hole-doped semiconductor wires. From Fig.~\ref{Fig1} we find that for electron-doped wires
 with a mobility of $100,000$ cm$^2$/V-s \cite{electronmobility},
which has been achieved for wires with a diameter $60$ nm, a measurable robust TS state gap of $50-100$ mK is achievable.
 From Fig.~\ref{Fig2} we find that for hole-doped wires with a
significantly larger spin-orbit energy $E_{SO}\sim 30$ meV, a larger gap of $0.8$ K is achievable for a
mobility of $100$ cm$^2$/V-s. \cite{holemobility} Thus, despite the fact that the semiconductor TS state breaks the
 time-reversal symmetry, realizing a robust TS state is possible in both electron- and hole-doped systems even in
 the presence of realistic disorder on the surface of the semiconductor.

Before concluding, it may be useful to provide a qualitative discussion and a general perspective on the issue of optimal
materials for the realization of the predicted TS phase and Majorana modes in semiconductor-superconductor sandwich
structures. The first point to emphasize is that there are far too many independent physical parameters entering
the problem for theory to be particularly useful in this context. At least six independent material parameters
play a role here: $\alpha$, $k_F$, $V_Z$, $\epsilon_{F,sm}$, $\Delta_s$, $t$ (which determines $\lambda$ in Eq.~\ref{lambda}).
In addition, disorder, which plays a crucial role by the non-existence of Anderson's theorem in the time-reversal invariance
broken (i.e. $V_Z\neq 0$) situation, is itself characterized by several independent physical mechanisms: disorder in the
adjacent bulk superconductor, long-ranged disorder in the semiconductor, disorder-induced fluctuations in $t$, interface
 disorder, puddles, and so on. A physical system
characterized by such a large ($>$ 10) set of independent parameters, all of which are important in determining the fate
of the TS state, is not an easy optimization problem to deal with theoretically. In particular, it may turn out to be
more practical to simply carry out experiments on available systems (e.g. InAs or InSb on Al or Nb) than to trust
precise theoretical quantitative predictions with respect to the optimal materials choice.

With these important caveats in mind, we can, nonetheless, make some general qualitative statements, quite apart from the
more precise quantitative statements made in the main body of this work. In the clean case, where disorder is not an issue
(i.e. extreme high-mobility semiconductor systems), it is obvious that the semiconductor spin-orbit coupling
should be as large as possible. This continues to be the case even in the presence of disorder, making $E_{SO} (\approx \alpha k_F)$
to be the key physical parameter, which should be made as large as possible (including enhancement by external gates if feasible ~\cite{Zhang-Xia})
with no constraints. Similarly, disorder is always harmful to the TS state, and therefore $1/\tau_{sm}$ should be made as small as
possible by working with the highest mobility materials. Although having a spin-orbit coupling strength as large as possible is obviously helpful to the realization of the TS phase, this may have the detrimental effect of enhancing impurity-induced scattering between the spin-split bands, thus reducing the mobility, \cite{Hwang} and therefore some optimization may be necessary even with respect to spin-orbit coupling and mobility in order to satisfy the general requirement of having both the highest possible spin-orbit coupling and mobility.

 The same is to a limited degree true of the superconducting gap $\Delta_s$
(or equivalently $T_c$) in the bulk supercnoductor, but this advantage is considerably nullified by the fact that the
proximity effect is determined primarily by the parameter $\lambda$ (see Eq.~\ref{delta0}), and this parameter should be made as large
as possible, even at the cost of having a lower $\Delta_s$ in the bulk superconductor (e.g. Al may be preferable over
Nb if $\lambda_{{\rm{Al}}}>\lambda_{{\rm {Nb}}}$). We note that having an arbitrarily large proximity gap $\Delta_0$ in the semiconductor may not
necessarily help unless $E_{SO}$ can also be simultaneously increased (e.g. by an external gate) since $E_{SO}>V_Z>2\lambda>2\Delta>2 E_g$ is a necessary condition for the existence of the TS phase. Similarly, having a very small $V_Z$ does not help
since $V_Z>2\Delta$ determines the condition for a robust topological phase, and having a small $V_Z$ necessarily requires
having a much smaller $\Delta$, which severely restricts the temperature range where the experiments investigating the
topological phase can occur since $T < \Delta$ is necessary. For the same reason a small $E_{SO}$, e.g. GaAs, with
correspondingly smaller $V_Z$ and $\Delta$ would not work either since the operational temperature would then
be too small.

As for disorder, our discussion in this work has considered the semiconductor disorder to be short-ranged, which is
consistent with interface disorder (or roughness) scattering which is known to be important in semiconductor
nanowires. Carrier screening should render long-range Coulomb disorder arising from random charged impurities
in the semiconductor into effective short-ranged disorder, making our model of rather general validity. The
fact that we model disorder through an effective relaxation time $\tau_{sm}$ defined through the measured carrier mobility makes
the model robust with respect to different types of disorder in the semiconductor, but we note that for long-ranged Coulomb
disorder $\tau_{sm}^{-1}$ is really only a lower bound on the effective disorder. (Disorder arising from impurities in the
superconductor does not affect the TS phase at all, see Ref.~\cite{Stanescu} for details.)

       A simple dimensional argument, which has been authenticated by a renormalization group
 calculation, \cite{Motrunich}
suggests $\Delta\tau_{sm}>1$, i.e. $\tau_{sm}^{-1}<\Delta$, as the bound on the effective
 scattering rate for the topological
gap $\Delta$ to survive disorder in a spinless chiral p-wave superconductor characterized only by these two parameters (unlike our problem which is characterized by many physical parameters). We note that our microscopic Born approximation calculation (Eq.~\ref{taut}) is consistent
with this case deep in the TS phase for $V_Z\sim E_{SO}$. For $E_{SO}\gg V_Z$, the effect of disorder scattering is suppressed
so that the topological gap survives disorder for a broader range of parameters $\Delta\tau_{sm}>\left({V_Z}/{E_{SO}}\right)^2$.
Unfortunately for the semiconductor-superconductor sandwich structures of interest to us, $\Delta\tau_{sm}>\left({V_Z}/{E_{SO}}\right)^2$ only provides a necessary condition since $\Delta$ cannot be made arbitrarily large unless $E_{SO}$ i.e. the spin-orbit coupling is itself arbitrarily
 large. The sufficient conditions for the gapped topological state to exist in the semiconductor sandwich structures
 are given by Eqs.~33, 34, which require ``large" $V_Z$ making it
necessary to have even larger $E_{SO}$!
Thus,  the existence of the TS phase in the realistic semiconductor sandwich structures depends on the seamless and multiparameter  materials optimization of  spin-orbit coupling, spin splitting, superconducting gap, and disorder in an inter-dependent manner.

We conclude with the observation that the single most important ingredient for the TS state to emerge in semiconductor-superconductor
 sandwich structures is to have a large spin-orbit coupling and high mobility. For electrons both InAs and InSb satisfy the
necessary requirements if the mobility can be enhanced around $10^5$ cm$^2$/V-s. For holes the spin orbit coupling being
very large, much lower mobilities ($\sim 10^2$ cm$^2$/V-s) should work. These considerations imply that
holes should be looked into experimentally for the realization of the TS phase. In addition, having several (odd number of)
subbands \cite{Lutchyn_prl} occupied in the nanowire should help since it reduces disorder through screening and enhances the effective values of
$E_{SO}$. We emphasize that the existence of the TS state depends delicately on the competition among superconducting proximity effect, spin-orbit 
coupling, spin splitting, and chemical potential.  (Disorder scattering always suppresses the TS state.)  As such, any materials optimization
 must carefully choose among these physical parameters while at the same time maintaining the highest possible mobility
 (i.e. smallest possible disorder scattering).
The experimental search for the Majorana mode should thus concentrate on semiconductor systems of the highest mobility and the strongest
 SO coupling using the appropriate superconductors which produce the best quality interface between the superconductor and the semiconductor.

 \section{acknowledgement}
 This work is supported by DARPA-MTO, DARPA-QuEST, JQI-NSF-PFC, Microsoft-Q, and NSF.
We thank Chuanwei Zhang for helpful discussions.

\appendix
\section{Dyson's equation and disorder averaged Green's function}\label{firstorderborn}
In this appendix, we start by first simplifying the Dyson equations Eqs.~\ref{dyson} and ~\ref{sigma} and
then solve the equations for the first order Born approximation case. The self-consistent Born approximation
is solved in Appendix.~\ref{selfconsistentborn}
Substituting Eqs.~\ref{Vmat} and~\ref{disordervertex} into Eq.~\ref{sigma}, the self-energy $\Sigma$ is written as
\begin{align}
&\Sigma(\bm k)=v^2\int d\bm k_1 [|\bra{\bm k}\ketn{\bm k_1}|^2\frac{(1+\tau_z)}{2}G(\bm k_1)\frac{(1+\tau_z)}{2}\nonumber\\
&+|\bra{-\bm k}\ketn{-\bm k_1}|^2\frac{(1-\tau_z)}{2}G(\bm k_1)\frac{(1-\tau_z)}{2}\nonumber\\
&-\bra{\bm k}\ketn{\bm k_1}\bra{-\bm k}\ketn{-\bm k_1}\frac{(1+\tau_z)}{2}G(\bm k_1)\frac{(1-\tau_z)}{2}\nonumber\\
&-\bra{\bm k}\ketn{\bm k_1}^*\bra{-\bm k}\ketn{-\bm k_1}^*\frac{(1-\tau_z)}{2}G(\bm k_1)\frac{(1+\tau_z)}{2}].\label{sigmad}
\end{align}
In principle, the disorder averaged scattering potential $\expect{V_{\alpha\lambda}(\bm k,\bm k_1)V_{\delta\beta}(\bm k_1,\bm k)}$
involves integrals of 4 Bloch eigenstates
 $\bra{\bm r,\sigma}\ketn{k}=u_{\bm k}(\bm r;\sigma)$
 of the form
 $v^2\int d\bm r \sum_{\sigma,\sigma'}u_{\bm k'}^*(\bm r;\sigma)u_{\bm k}(\bm r;\sigma)u_{\bm k}^*(\bm r;\sigma')u_{\bm k'}(\bm r;\sigma')$
 where the sum over $\sigma$ implies that the
disorder is spatially local spin-independent disorder.
Having several sub-bands occupied in the 1D nanowire, as has recently been proposed \cite{Stanescu,Lutchyn_prl}, should help
the situation considerably, both by expanding the regime of stability of the TS phase and by screening out the disorder
due to higher carrier density.
Substituting the wave-function into the matrix elements,
 we get
\begin{align}
&\bra{|k|,\theta_{\bm k}}\ketn{|k'|,\theta_{\bm k'}}\approx \bra{k_F,\theta_{\bm k}}\ketn{k_F,\theta_{\bm k'}}\nonumber\\
&=\bra{0}e^{i(R_z-\lambda)(\theta_{\bm k}-\theta_{\bm k'})}\ket{0}=f(\theta_{\bm k}-\theta_{\bm k'}).
\end{align}
Substituting the matrix elements into the disorder-induced self-energy we find that,
\begin{align}
&\Sigma(\bm k,\omega)=v^2\int d\bm k_1 [|f(\theta_{\bm k}-\theta_{\bm k_1})|^2\frac{(G(\bm k_1)+\tau_z G(\bm k_1)\tau_z)}{2}\nonumber\\
&-f(\theta_{\bm k}-\theta_{\bm k_1})^2\frac{(1+\tau_z)}{2}G(\bm k_1)\frac{(1-\tau_z)}{2}\nonumber\\
&-f(\theta_{\bm k}-\theta_{\bm k_1})^{* 2}\frac{(1-\tau_z)}{2}G(\bm k_1)\frac{(1+\tau_z)}{2}].\label{sigma1}
\end{align}
To solve Dyson's equations, it is convenient to define the functions $a_{0,z,+,-}(\bm k,\omega)$ by the relation
\begin{equation}
G^{-1}(\bm k,\omega)=a_0(\bm k,\omega)+a_z(\bm k,\omega)\tau_z+a_+(\bm k,\omega)\tau_++a_-(\bm k,\omega)\tau_-.\label{ginv}
\end{equation}
Substituting this into the Dyson equations (Eq.~\ref{dyson}) together with the expression for the self-energy (Eq.~\ref{sigma1})
in the lowest order Born approximation  we get
\begin{align}
&a_0(\bm k,\omega)=\omega-f_0 (\pi \tau_{sm})^{-1}\int d\epsilon' \frac{\omega}{\omega^2-(\epsilon'-\epsilon_{F,sm})^2-|\Delta|^2}\nonumber\\
&a_z(\bm k,\omega)=(\epsilon-\epsilon_{F,sm})+f_0 (\pi \tau_{sm})^{-1}\int d\epsilon' \frac{(\epsilon'-\epsilon_{F,sm})}{\omega^2-(\epsilon'-\epsilon_{F,sm})^2-|\Delta|^2}\nonumber\\
&a_+(\bm k,\omega)=\Delta  e^{i\theta_{\bm k}}-f_1 (\pi \tau_{sm})^{-1}\int d\epsilon' \frac{\Delta  e^{i\theta_{\bm k}}}{\omega^2-(\epsilon'-\epsilon_{F,sm})^2-|\Delta|^2},\label{borneq}
\end{align}
where $f_0=\int \frac{d\theta}{2\pi}|f(\theta)|^2$, $f_1=\int \frac{d\theta}{2\pi}f(\theta)^2 e^{i\theta}$.
The integrals in Eq.~\ref{borneq} can be written analytically as
\begin{align}
&a_0=\omega+f_0 (\pi\tau_{sm})^{-1}\zeta\omega\\
&a_+=[\Delta+f_1 (\pi\tau_{sm})^{-1}\zeta\Delta]e^{i\theta_{\bm k}}
\end{align}
where
\begin{equation}
\zeta=\int d\epsilon \frac{1}{\epsilon^2+\Delta^2-\omega^2}=\frac{\pi}{\sqrt{\Delta^2-\omega^2}}
\end{equation}
and $a_z(\bm k,\omega)=(\epsilon_{\bm k}-\epsilon_{F,sm})$ remains unrenormalized.

The lowest-frequency pole in the disorder averaged Green function $G(\bm k,\omega)$
 given by Eq.~\ref{ginv}, which
corresponds to the quasiparticle gap, occurs when
 $\epsilon_{\bm k}-\epsilon_{F,sm}=a_z=0$ and $a_0=|a_+|$. The disorder averaged
quasiparticle gap $\omega$
can then be obtained by solving
\begin{equation}
\omega[1+\frac{f_0}{\sqrt{\Delta^2-\omega^2}}]=\Delta[1+\frac{f_1}{\sqrt{\Delta^2-\omega^2}}].
\end{equation}
For sufficiently weak disorder $\tau_{sm}^{-1}\ll\Delta$, the quasiparticle gap
 $\omega\sim \Delta$, so that we can expand in
the reduction of the quasiparticle gap $x=\Delta-\omega$ to lowest order
and obtain the quasiparticle gap reduction to be
\begin{equation}
x\sim \frac{\Delta}{2^{1/3}}\left(\frac{(\pi\tau_{sm})^{-1}}{\Delta}\right)^{2/3}\left(f_0-f_1\right)^{2/3}.
\end{equation}
This leads to the pole of the disordered average Green function $G$ which is also the TS state quasiparticle gap
in Eq.~\ref{eq:x}
\begin{equation}
\omega=\Delta\left[1-\frac{1}{2^{1/3}}\left(\pi \tau_{sm}\Delta\right)^{-2/3}|\expect{k_F|-k_F}|^{4/3}\right].
\end{equation}

\section{Analytic solution of the self-consistent Born approximation}\label{selfconsistentborn}
In this appendix we provide an exact solution for the quasiparticle gap within the self-consistent
Born approximation. To do this, we start with the self-consistent version of Eq.~\ref{borneq},
\begin{align}
&a_0(\bm k,\omega)\nonumber\\
&=\omega\nonumber\\
&-f_0 (\pi\tau_{sm})^{-1}\int d\epsilon' \frac{a_0(\bm k,\omega)}{a_0(\bm k,\omega)^2-(\epsilon'-a_z(\bm k,\omega))^2-|a_+(\bm k,\omega)|^2}\nonumber\\
&a_z(\bm k,\omega)=(\epsilon-\epsilon_{F,sm})\nonumber\\
&+f_0 (\pi\tau_{sm})^{-1}\int d\epsilon' \frac{(\epsilon'-\epsilon_{F,sm})}{a_0(\bm k,\omega)^2-(\epsilon'-a_z(\bm k,\omega))^2-|a_+(\bm k,\omega)|^2}\nonumber\\
&a_+(\bm k,\omega)=\Delta  e^{i\theta_{\bm k}}\nonumber\\
&-f_1 (\pi\tau_{sm})^{-1}\int d\epsilon' \frac{a_+(\bm k,\omega)}{a_0(\bm k,\omega)^2-(\epsilon'-a_z(\bm k,\omega))^2-|a_+(\bm k,\omega)|^2}.\label{scborneq}
\end{align}
Performing the $\epsilon'$ integral in the second equation we immediately obtain $a_z(\bm k,\omega)=0$. Performing the $\epsilon'$
integral in the other 2 equations
\begin{align}
&a_0=\omega+a\frac{a_0}{\sqrt{a_+^2-a_0^2}}\\
&a_+=\Delta+b\frac{a_+}{\sqrt{a_+^2-a_0^2}}
\end{align}
where the square root is chosen so that the real part of $\sqrt{a_+^2-a_0^2}$, $a=\pi f_0 (\pi\tau_{sm})^{-1}$
 and $b=\pi f_1 (\pi\tau_{sm})^{-1}$. Writing $y=\sqrt{a_+^2-a_0^2}$, the pair of
equations can be combined to a single equation for $y$
\begin{equation}
1=\frac{\Delta^2}{(y-b)^2}-\frac{\omega^2}{(y-a)^2}.\label{eqy}
\end{equation}
At $\omega\sim 0$, the equation has 4 real roots $y\sim a$ and $y\sim b\pm\Delta$.
The quasiparticle gap is defined by the value of $\omega$ above which the Green function $G(\omega)$ has
a finite imaginary part. This corresponds to the value of $\omega$ where 2 of the real roots of $y$ merge
into a pair of complex roots, so that Eq.~\ref{eqy} has 2 real and 2 complex roots. Using the
properties of quartic polynomial equations, the transition point between 4 real roots and 2 real roots,
is signalled by the vanishing of the discriminant of the quartic polynomial. Solving for the frequency $\omega$
where the discriminant vanishes leads to the solution
\begin{equation}
\omega=\Delta\sqrt{1-3 g^{2/3}(f_0-f_1)^{2/3}+3g^{4/3}(f_0-f_1)^{4/3}-3g^2(f_0-f_1)^2}
\end{equation}
where $g=\frac{1}{\pi\Delta\tau_{sm}}$.
In the 2D Rashba and 1D superconducting case, we get the solution of the TS state quasiparticle gap in
Eq.~\ref{scbornsol}
\begin{align}
&\omega=\Delta[1-3\left(\pi\tau_{sm}\Delta\right)^{-2/3}|\expect{k_F|-k_F}|^{4/3}\nonumber\\
&+3\left(\pi\tau_{sm}\Delta\right)^{-4/3}|\expect{k_F|-k_F}|^{8/3}-3\left(\pi\tau_{sm}\Delta\right)^{-2}|\expect{k_F|-k_F}|^{4}]^{1/2}.
\end{align}





\end{document}